\documentstyle[psfig]{l-aa}
\begin{document}
\renewcommand{\topfraction}{1.}
\renewcommand{\bottomfraction}{1.}
\renewcommand{\textfraction}{0.}
\thesaurus{06(02.13.3; 08.03.4; 08.13.2; 08.19.3; 11.13.1; 13.19.5)}
\title{Discovery of the first extra-galactic SiO maser\thanks{based on
       observations obtained at the European Southern Observatory}}
\author{Jacco Th. van Loon\inst{1,2}, Albert A. Zijlstra\inst{1},
        Valent\'{\i}n Bujarrabal\inst{3} \and Lars-{\AA}ke
        Nyman\inst{4,5}}
\institute{European Southern Observatory, Karl-Schwarzschild
           Stra{\ss}e 2, D-85748 Garching bei M\"{u}nchen, Germany
      \and Astronomical Institute ``Anton Pannekoek'', University of
           Amsterdam, Kruislaan 403, NL-1098 SJ Amsterdam,
           The Netherlands
      \and Observatorio Astron\'{o}mico Nacional, Campus
           Universitario, Apartado 1143, E-28800 Alcala de Henares
           (Madrid), Spain
      \and European Southern Observatory, Casilla 19001, Santiago 19,
           Chile
      \and Onsala Space Observatory, S-439 92 Onsala, Sweden}
\date{Received date; accepted date}
\maketitle
\markboth{Jacco Th.\ van Loon et al.: Discovery of the first
          extra-galactic SiO maser}{Jacco Th.\ van Loon et al.:
          Discovery of the first extra-galactic SiO maser} 
\begin{abstract}

We report on the detection of SiO J=2--1 v=1 maser emission from
the red supergiant IRAS04553--6825 in the LMC. It has thereby become
the first known source of SiO maser emission outside the Milky Way.

\end{abstract}

\section{Introduction}

The heaviest post-main sequence mass-loss occurs during the Red
Supergiant (RSG, high mass stars) phase or at the tip of the
Asymptotic Giant Branch (AGB, intermediate mass stars). In oxygen-rich
environments, the velocity structure of the inner part of the
circumstellar envelope below the dust formation can be studied in SiO
maser emission (Chapman \& Cohen 1986). This is where the poorly
understood initiation of the mass loss flow occurs.

Recently, the first obscured evolved stars in the LMC were found,
using the IRAS database (Reid 1991; Wood et al.\ 1992; Zijlstra et
al.\ 1995). Wood et al.\ (1992) discovered OH maser emission from
six of these stars: the first known extra-galactic OH/IR stars. We
used the SEST at La Silla to observe the best candidates for detection
of 86~GHz SiO maser emission among these OH/IR stars. As a first
result, we present the discovery of the first extra-galactic SiO maser
--- viz. the RSG IRAS04553--6825.

\section{Observations}

The observations were performed on days 25, 30, and 31 of May 1995.
We used the 15m SEST (sub)mm telescope at ESO/La~Silla with the new
3~mm SIS receiver as the frontend, and the high resolution
spectrograph HRS as the backend. We observed the vibrationally excited
rotational transition of the silicate molecule SiO(2--1)$_{v=1}$ at
86243.442~MHz. During part of the time we used this configuration
simultaneously in combination with the low resolution spectrograph LRS
tuned at the same frequency. We observed in a dual beam switch mode
with the source alternately placed in the two beams, a method which
yields very flat baselines. The beam separation was $\sim
11.5^{\prime}$. The pointing was checked every few hours, using the
nearby bright SiO maser R~Dor. This agreed with the specified accuracy
of $3^{\prime\prime}$ rms. The Full Width Half Power of the beam is
$57^{\prime\prime}$. At 86~GHz the channel separation is
0.15~km~s$^{-1}$ for the HRS, and 2.4~km~s$^{-1}$ for the LRS. The
conversion factor from antenna temperature to flux units is
25~Jy~K$^{-1}$ at 86~GHz. The internal absolute flux calibration is
accurate to about 20\%.

Helped by good atmospheric conditions --- relative humidity 15 -- 30\%,
and outside air temperature about 15$^\circ$C with little or no cirrus
--- the system temperature at 86~GHz was about 120~K or less at
elevations above 40$^\circ$, and about 150~K at an elevation of
20$^\circ$. Polarization biases were minimized by observing at
different hour angles, because the parallactic angle changes with hour
angle. Long integrations naturally take care of this.

\section{Detection criteria}

With the spectrograph having 1600 -- 2000 channels --- though not all
independent of each other --- a minimum value for a detection
threshold is three times the rms noise level. However, the
significance of a resolved spectral feature is higher than its peak
intensity suggests.

A better estimate of the significance of a resolved spectral feature
is obtained by rebinning the spectrum. By dividing this smoothed
spectrum by its rms noise level, the peak of the spectral feature is
expressed in terms of rms noise level. The peak of the spectral
feature will reach the highest significance for one specific binning.
This yields the spectral width of the feature and the significance of
the integrated spectral feature.

An additional check can be made on a possibly detected maser peak.
Since the spectra are averages of many individual scans, statistics
can be done on this series of scans. For a specific channel we can
calculate the variance over the set of scans, yielding an estimate
for the noise level of each channel. A feature in the variance
spectrum coincident with a possibly detected maser peak casts doubt on
the maser nature of the peak. A maser peak is not expected to vary
intrinsically within the course of an observation, which makes an
anomaly in the noise more likely. The variance spectrum can also be
used to check the derived average rms noise level. In the same way, we
can calculate the covariance between one channel and another channel
over the set of scans. For each channel, we can integrate this
covariance over a fixed number of adjacent channels, obtaining the
covariance spectrum. Again, a feature in the covariance spectrum right
at the position of the possibly detected maser peak would suggest an
anomaly in the noise instead.

We used a few different frequency settings during the course of the
observation to check for systematic flux offsets due to bad channels.
The distribution of the observations over three separate shifts allows
for a check on consistency too. The very unlikely case of an
artificial signal in the HRS spectrograph can be ruled out by
simultaneous observation with the LRS spectrograph. Finally, proximity
in velocity space to an OH detected maser feature further enhances the
significance of a possible SiO detection.

\section{Results}

%
%
\begin{figure}[htb]
\centerline{\vbox{
\psfig{figure=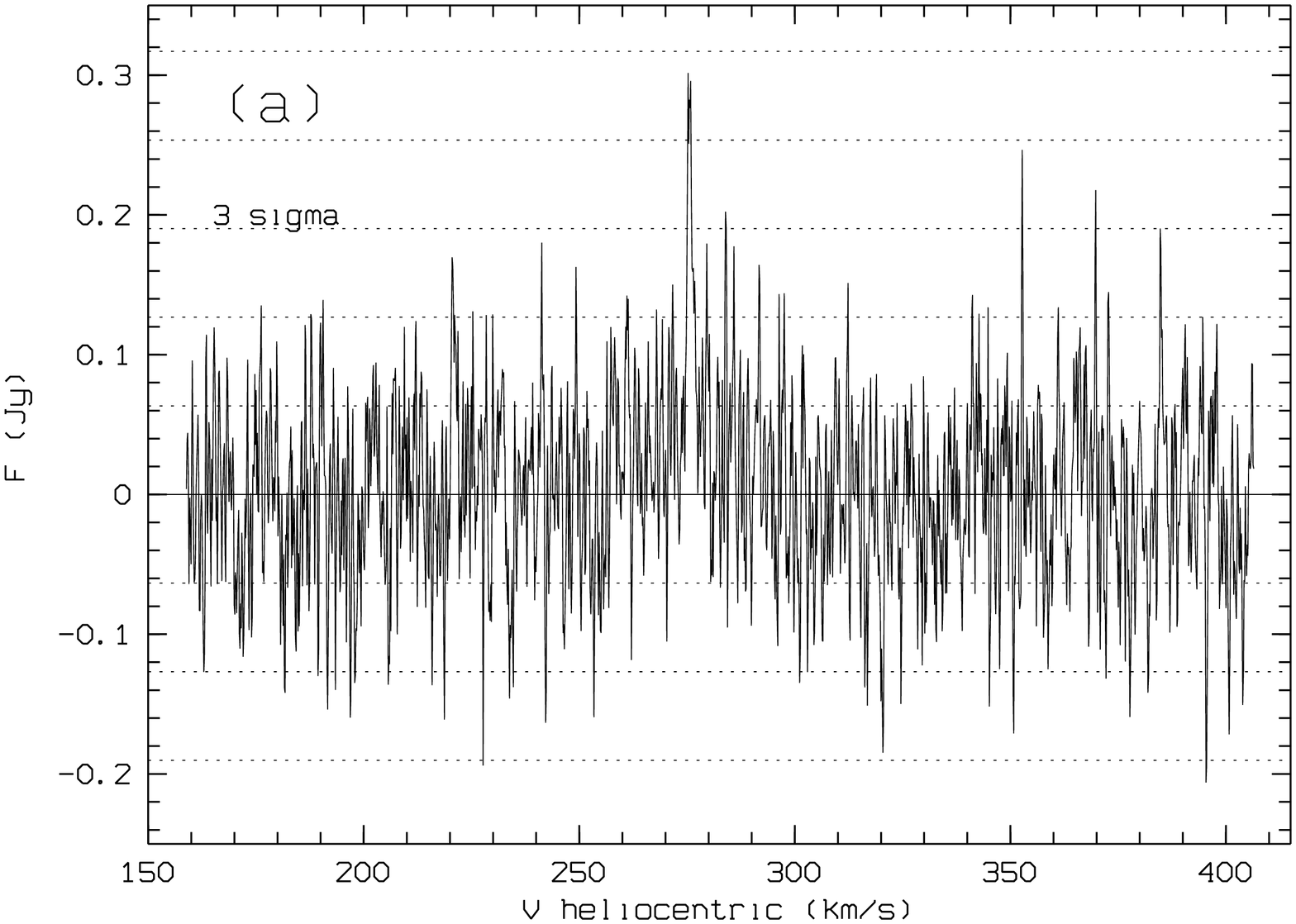,width=95mm,height=64mm,angle=270}
\psfig{figure=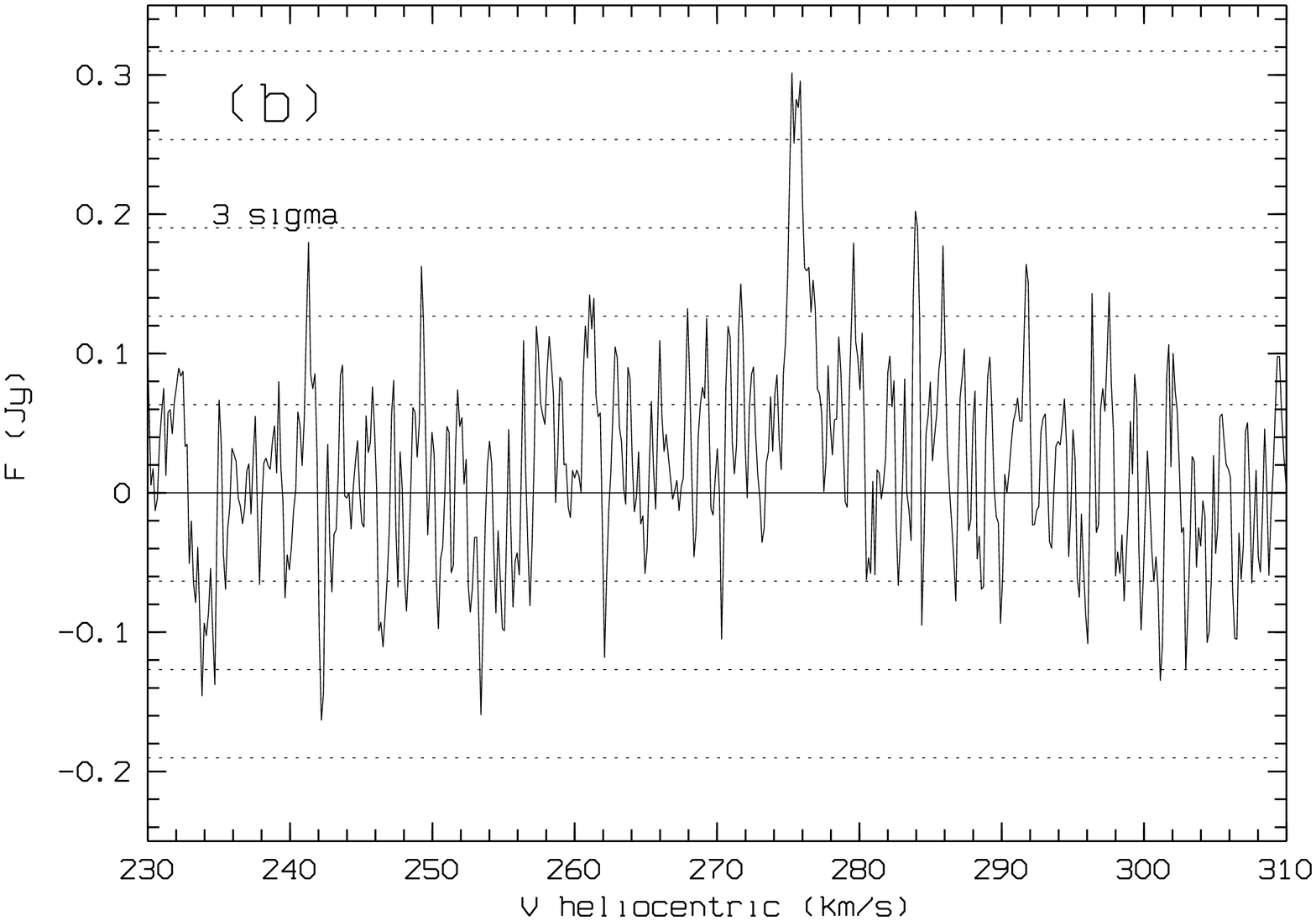,width=95mm,height=64mm,angle=270}
\psfig{figure=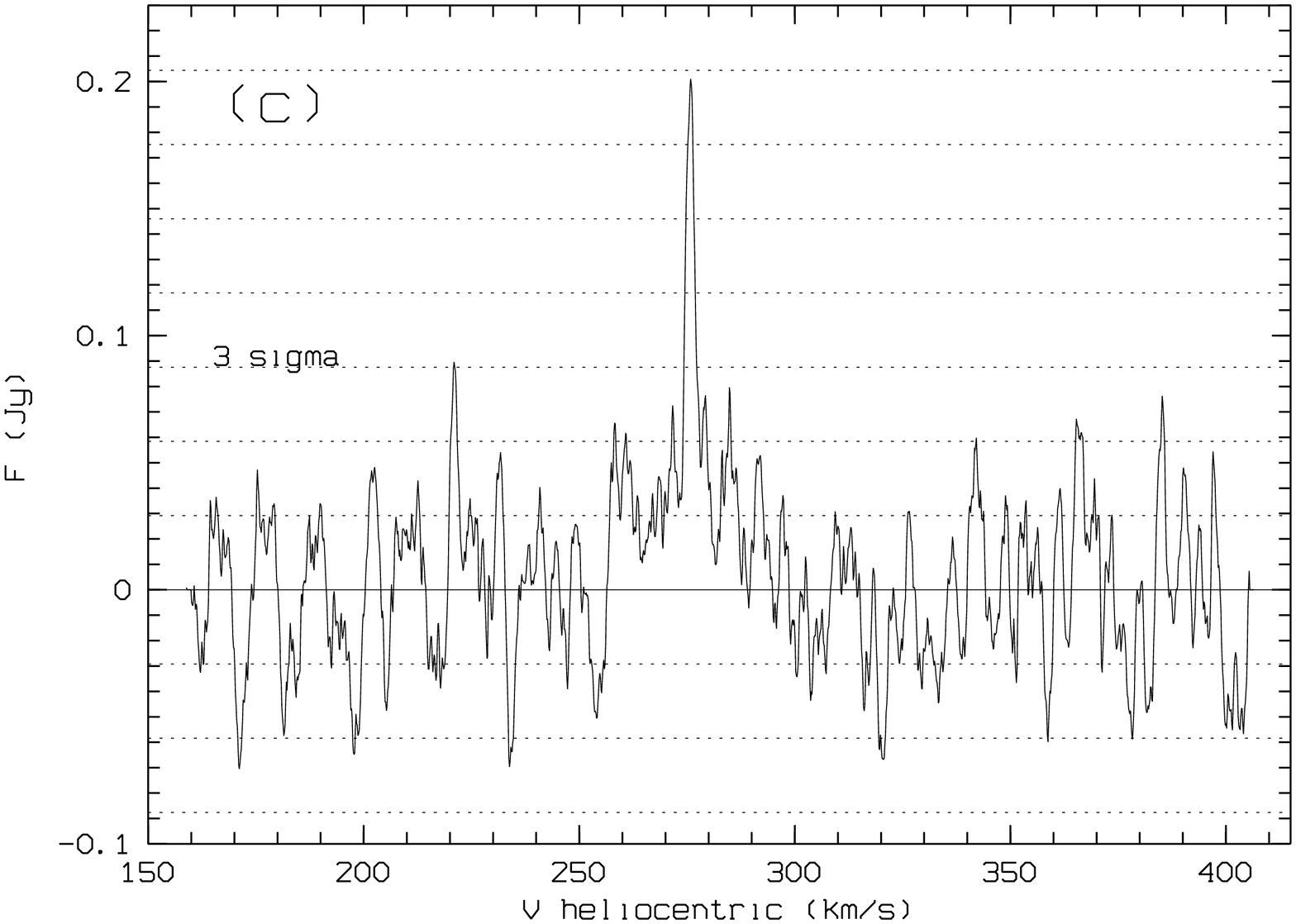,width=95mm,height=64mm,angle=270}
}}
\caption[]{(a) High resolution (HRS) spectrum around the
SiO(2--1)$_{v=1}$ maser emission from IRAS04553--6825. The velocities
are heliocentric, and horizontal dotted lines are given per
1~${\sigma}$ (1~${\sigma}$ = 63~mJy). (b) Expanded section of the
original HRS spectrum around the maser peak. (c) The HRS spectrum
smoothed by averaging over 15 channels (= 2.25~km~s$^{-1}$). Now
1~${\sigma}$ = 29~mJy}
\end{figure}

All HRS spectra of the SiO(2--1)$_{v=1}$ transition of IRAS04553--6825
are added together. The resulting spectrum is shown in Fig.\ 1 (a and
b). This spectrum is the result of a total of 9.9 hours on-source
integration time. Due to the beamswitching and overhead from
reading/writing, tuning, and pointing checks, the actual observing
time was $\sim 26$ hours. The 1~${\sigma}$ level (= rms) is only
63~mJy. This low a noise level could only be achieved by the excellent
performance of the new receiver.

The SiO maser emission is clearly seen as a narrow, yet resolved peak
at v$_{\rm hel} \sim 275 - 276$~km~s$^{-1}$, at a level of
$\sim 4.6$~${\sigma}$. We replaced the flux value of each channel by
the average flux within a bin, centered on that channel and 15
channels wide (= 2.25~km~s$^{-1}$). The spectrum is shown in Fig.\ 1
(c). Now, the significance of the maser is 6.9~${\sigma}$ (1~${\sigma}
= 29$~mJy). The maser emission peaks at v$_{\rm hel} = 275.6 \pm
0.3$~km~s$^{-1}$, with a peak flux F$_{\rm max} = 0.28 \pm 0.06$~Jy
(Fig.\ 1 (b)). The FWHM of the feature is $\sim 1.1$~km~s$^{-1}$. The
total width of the peak is $\sim 2.3$~km~s$^{-1}$, with the
center-of-gravity at v$_{\rm hel} \sim 275.9$~km~s$^{-1}$. Integrating
the feature in the region v$_{\rm hel} = 274.73 - 276.99$~km~s$^{-1}$,
we obtain an integrated flux F$_{\rm int} = 0.45 \pm
0.07$~Jy~km~s$^{-1}$. There is a hint of emission at 260~km~s$^{-1}$,
as well as a possible broad pedestal emission component at v$_{\rm
hel} \sim 260 - 290$~km~s$^{-1}$. If real, this would imply an
integrated flux of about a factor two higher than estimated above.

The variance and covariance spectra were featureless. The peak is
also dominantly present in the LRS spectrum that we took during 5.7
hours out of the total on-source integration time. Thus, the
maser peak satisfies all the above-mentioned detection criteria.

\section{Discussion}

The optical counterpart of IRAS04553--6825 is the red supergiant
WOH~G064 (Westerlund et al.\ 1981; Elias et al.\ 1986). It is
extremely bright, both bolometric (M$_{\rm bol} \sim -9.3$, Zijlstra
et al.\ 1995) as well as in the mid-IR (F$_{12 {\mu}m} \sim 8$~Jy).
The progenitor mass is probably as massive as $\sim 50 M_{\sun}$. For
this massive a star, the spectral type is very late: about M7. The
severe circumstellar extinction estimated from the (probably
self-absorbed) 9.7~$\mu$m silicate feature seems to be inconsistent
with the relatively moderate circumstellar extinction estimated from
the optical. This has usually been taken as an indication for either
binarity or a highly flattened circumstellar material distribution
(Roche et al.\ 1993).

SiO maser emission intensity is correlated with the mid-IR luminosity
(e.g.\ Jewell et al.\ 1991; Lane 1982; Nyman et al.\ 1993). OH/IR
stars which satisfy K -- L$ < 4$ mag exhibit brighter
SiO(2--1)$_{v=1}$ masers than redder stars (Nyman et al.\ 1993). The
SiO maser emission intensity can be highly variable in time, but
typically peaks in the first quarter of the pulsation cycle (Lane
1982; Nyman \& Olofsson 1986). At the time of the observations
IRAS04553--6825 satisfied all of these criteria for being an
outstanding candidate for detection of SiO(2--1)$_{v=1}$ maser
emission.

IRAS04553--6825 is the first extra-galactic stellar OH maser discovered
(Wood et al.\ 1986) and the strongest stellar 1612~MHz OH maser source
in the LMC; it also exhibits 1665~MHz mainline OH maser emission
(Wood et al.\ 1992). Both OH masers are double peaked, from which Wood
et al.\ derived a stellar velocity of v$_{\star}$(OH)$ \sim
260$~km~s$^{-1}$. The bulk of the presently detected SiO maser
emission comes from a peak that is situated redward of the OH maser
emission, with a velocity difference with respect to v$_{\star}$(OH)
of v$_{\rm hel} \sim 16$~km~s$^{-1}$. This is atypical, since SiO
maser emission is usually observed to be centered at the stellar
velocity (e.g. Heske 1989; Jewell et al.\ 1991; Lewis et al.\ 1995;
Nyman et al.\ 1986). SiO maser emission is known to be highly variable
in both intensity and line shape in an erratic way on timescales of
months (Lane 1982), but excursions of SiO maser peaks only reach
several km~s$^{-1}$ with respect to the stellar velocity (e.g.\
Bujarrabal et al.\ 1986; Lane 1982; Nyman \& Olofsson 1986; Olofsson
et al.\ 1985). Thus, it seems likely that the OH is not centred on the
stellar velocity. An indication that not all the OH emission has been
observed comes from the very low value for the expansion velocity
derived from the OH of only 11~km~s$^{-1}$: this is a factor of
3 -- 4 lower than is normally found for supergiants. To confirm this
hypothesis, we took a medium resolution spectrum (R $\sim$ 75,000)
with the NTT (ESO) in October 1995: the spectrum shows H$\alpha$
emission peaking at v$_{\rm hel} \sim 274$~km~s$^{-1}$ (extending
between v$_{\rm hel} \sim 265 - 293$~km~s$^{-1}$), confirming the
revised value for the stellar velocity.

The difference between the SiO maser peak velocity and the blue-most
edge of the OH emission at v$_{\rm hel} \sim 250$~km~s$^{-1}$ yields
an expansion velocity of v$_{\rm exp} \sim 26$~km~s$^{-1}$, much more
compatible with Milky Way RSGs than the exceptionally low outflow
velocity of v$_{\rm exp} \sim 11$~km~s$^{-1}$ previously determined.
The red-shifted OH emission is not observed. The latter is expected if
there is an inner ionized region which is optically thick at 18~cm
(e.g.\ Shepherd et al.\ 1990). The spectral type of IRAS04553--6825 is
not consistent with an ionized stellar wind. The discovery of
forbidden [NII] lines (Elias et al.\ 1986) confirms the existence of
ionized gas in its vicinity, but it is not clear whether this HII
region is centred on the star.

From the integrated flux of the maser emission, and assuming that the
maser emits isotropically, we derive a total photon flux of $\sim 6.8
\times 10^{44}$ photons/s. SiO maser emission is thought to be
tangentially amplified (Diamond et al.\ 1994). For maser emission
coming from a rather discrete structure such as a globule, the angle
between the maser beam and the line of sight can be non-zero. In
that case, the assumption of isotropic emission is no longer valid,
which may lead to either over- or under-estimating the total photon
flux. Assuming isotropic emission, the observed total photon flux
places IRAS04553--6825 at the bright end of the distribution of the
Milky Way AGB stars. The total photon fluxes of the RSGs VY~CMa and
VX~Sgr are an order of magnitude larger (Lane 1982), but they are
exceptionally bright. Numerical simulations yield larger total photon
fluxes with increasing stellar radius (Bujarrabal 1994). Observations
showed that the total photon flux also increases with increasing
pulsation period (Lane 1982). In this respect, IRAS04553--6825 may not
be that similar to Milky Way sources but rather underluminous, since
from its very large radius (it is very bright and very cool) and its
relatively long period (P~$\sim 930$~d) we might have expected
IRAS04553--6825 to be at least as bright as VY~CMa and VX~Sgr.

Alcolea et al.\ (1990) found that Mira variables always have a ratio
SiO/IR (= SiO(2--1)$_{v=1}$ peak intensity/8~$\mu$m intensity) $\sim
1/10$ (see also Hall et al.\ 1990). This is the maximum pump
efficiency and it is only reached in favorable conditions (Bujarrabal
et al.\ 1987). Semi Regular Variables (SRVs)s with visual amplitudes
exceeding 2.5 magnitude always have about the maximum efficiency,
while SRVs with lower amplitude in V rarely reach the maximum
efficiency and are often more than an order of magnitude less
efficient. VY~CMa and VX~Sgr have ratios SiO/IR~$\sim 1/7$ and 1/4,
and large amplitudes in V of resp.\ 3.1 and 7.5 mag. This may be (part
of) the cause why these two RSGs exhibit such bright SiO maser
emission. IRAS04553--6825 has SiO/IR~$\sim 1/30$, which means that it
is quite efficient, but not optimal. The small amplitude of 0.3
magnitude in K suggests that IRAS04553--6825 is not a large amplitude
variable, although the amplitude in V is unknown. IRAS04553--6825 has
an integrated- over peak flux ratio of F$_{\rm int}$/F$_{\rm max} \sim
2$~km~s$^{-1}$, whereas all detections in Alcolea et al.\ 1990 have
F$_{\rm int}$/F$_{\rm max} = 2.5 - 6.1$~km~s$^{-1}$ (Miras) and F$_{\rm
int}$/F$_{\rm max} \sim 7.0 - 10.7$~km~s$^{-1}$ (SRVs). This is another
indication that IRAS04553--6825 is probably underluminous in terms of
total photon flux.

%
%
\begin{figure}[htb]
\centerline{\psfig{figure=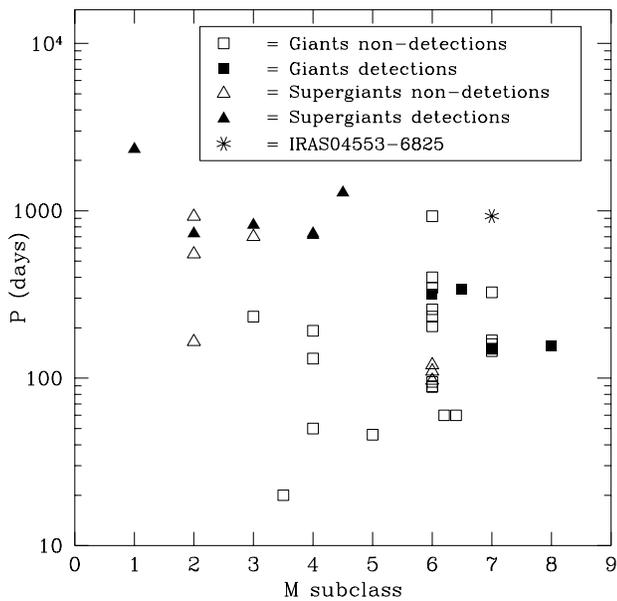,width=85mm}}
\caption[]{Periods and spectral types of giants (squares) and RSGs
(triangles) from which Alcolea et al.\ 1990 did (solid symbols) or did
not (open symbols) detect SiO(2--1)$_{v=1}$ emission. The star
represents IRAS04553--6825}
\end{figure}

In the sample of Alcolea et al.\ (1990), all stars with periods
exceeding 400 days have spectral types M6 or earlier. The RSGs with
detected SiO(2--1)$_{v=1}$ maser emission all have spectral type M4.5
or earlier, whereas detected giants have later spectral types but
shorter periods (solid symbols in Fig.\ 2). The M4.5 RSG NML~Cyg
(pulsation period P~$\sim 1280$~d) may be the Milky Way star with the
closest resemblance to IRAS04553--6825. Alcolea et al.\ detected
SiO(2--1)$_{v=1}$ maser emission from this star with a pump efficiency
of only SiO/IR~$\sim 1/190$. However, the total photon flux (relative
to the 8~$\mu$m intensity) is similar to that from IRAS04553--6825.
The same is true for their energy distributions (Elias et al.\ 1986).

\section{Conclusions}
We discovered the first extra-galactic SiO maser, from the red
supergiant IRAS04553--6825 in the LMC. The SiO maser peak was situated
16~km~s$^{-1}$ redward of the center of the double peaked OH maser
emission. We argue that the SiO maser peak velocity coincides with the
stellar velocity. This would mean that the outflow velocity of the
circumstellar matter around IRAS04553--6825 is v$_{\rm exp} \sim
26$~km~s$^{-1}$, which is typical for galactic RSGs. The peak
intensity of the SiO maser emission is not incompatible with ranges
found in galactic RSGs, but the total integrated photon flux is lower
than expected.

\acknowledgements{We thank Peter te Lintel Hekkert and Rens Waters for
helpful discussion, and Lex Kaper and Pascal Ballester for help with
the preliminary analysis of the optical spectrum. Jacco agradece sobre
todo a Montse, much\'{\i}simo.}

\end{document}